# Multiphasic profiles for the ion activities of NaBr and KBr

Per Nissen

Norwegian University of Life Sciences
Department of Ecology and Natural Resource Management
P. O. Box 5003, NO-1432 Ås, Norway

per.nissen@nmbu.no

# Introduction

The preceding paper (Nissen 2017) shows multiphasic profiles, straight lines separated by discontinuous transitions, for the ion activities of NaCl and KCl at 15, 25, 35 and 45ºC based on very precise tabular data of Lee et al. (2002). In addition, references are given to previous findings of multiphasic profiles for a variety of processes and phenomena, starting with the uptake of sulfate by roots and leaf slices of barley (Nissen 1971).

In a continuation, the present paper shows multiphasic profiles for the ion activities of NaBr and KBr for data also from Lee et al. (2002).

# Reanalysis

Data for the activity coefficients of Na$^+$ and Br$^-$ in aqueous solutions of various concentrations at 288.15, 298.15, 308.15 and 318.15 K (Tables 3 and 7 in Lee et al. 2002) have been plotted against each other (Figs 1-8). Data for the activity coefficients of K$^+$ and Br$^-$ (Tables 4 and 8) have also been plotted (Figs 9-16).

As before (Nissen 2017), the profiles are very well represented by straight lines (54 of the 58 absolute r values in Figs 1-16 are 0.999 or higher). The transitions are clearly discontinuous when the lines intersect in a common point (between lines VIII and IX in Fig. 2, between lines V, VI, VII and VIII in Fig. 3, between lines I and II in Figs 5 and 6, between lines II and III in Fig. 11). Transitions close to a point are probably also discontinuous.

When adjacent lines are parallel or nearly so, the transition is necessarily in the form of a jump, as is the case for the profile for Na$^+$ (lines II and III, and lines VII and VIII) at the lowest temperature (Fig. 1). There are similar findings also for Figs 2-4, showing that the phases are virtually unaffected by temperature, with no or only minor changes in the pattern. The profiles for Br$^-$ decrease with increasing NaBr concentrations, but there may be slight and temporary increases at intermediate and high concentrations.

The profiles for K$^+$ decrease consistently with increasing concentrations (Figs 9-12). The profiles for Br$^-$ decrease at first with increasing KBr concentrations, but then increase somewhat (Figs 13-16). Lines II-III are parallel at all temperatures and are separated by a tiny jump.

The description of this reanalysis is somewhat less complete than in the previous paper (Nissen 2017), but the figures and their legends provide compelling evidence that the profiles are indeed multiphasic and cannot be validly represented as curvilinear. The present profiles for Na$^+$ (Figs 1-4) are especially complex, consisting as they do of four sections: A rapid initial decrease at low concentrations, an increase followed by a decrease at intermediate concentrations, and a final rapid increase at high concentrations. However, the data are still represented as curvilinear by the authors (Fig. 4, upper profile, in Lee et al. 2002) using a modified Pitzer-Debye-Hückel model. This is made possible by their unmentioned omission of the four circled points in Fig. 17. When these points are included the profile is clearly multiphasic.

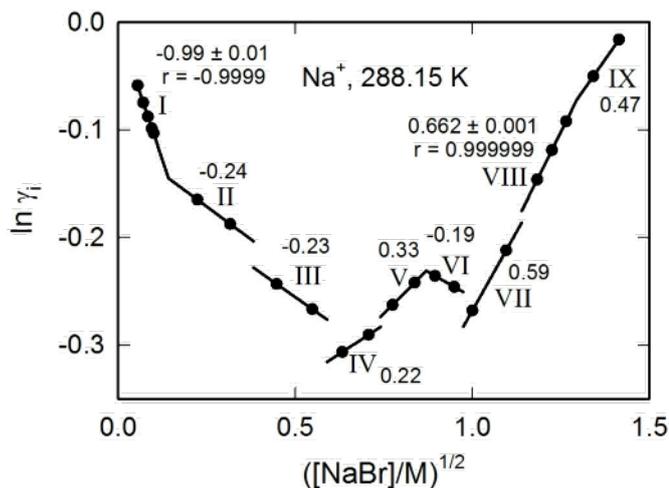

**Fig. 1.** Nine phases. Transitions at 0.142, between 0.316 and 0.447 (jump), between 0.548 and 0.632 (jump), and between 0.707 and 0.775 (small jump), at 0.870, between 0.949 and 1.000 (jump), between 1.095 and 1.183 (small jump), and at 1.296. Lines II and III are parallel, as are approximately also lines VII and VIII. The absolute r value of line I is very high, that of line VIII is exceedingly high.

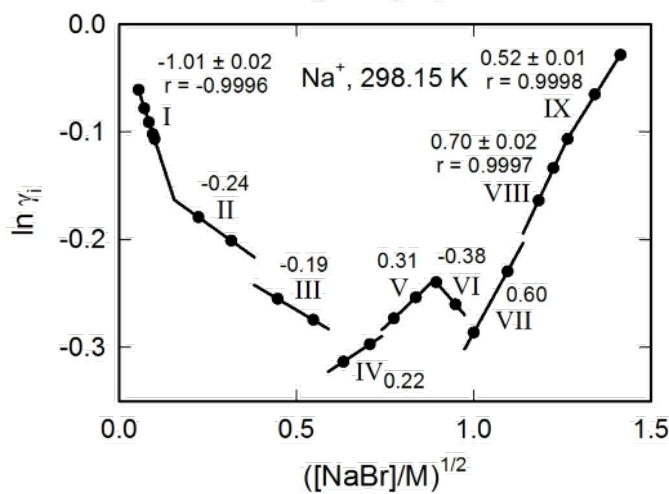

**Fig. 2.** Nine phases. Transitions at 0.155, between 0.316 and 0.447 (jump), between 0.548 and 0.632 (jump), and between 0.707 and 0.775 (small jump), at 0.889, between 0.949 and 1.000 (jump), between 1.095 and 1.183 (small jump), and at 1.265. Lines II and III are about parallel, as are also lines VII and VIII. The absolute r values of lines I, VIII and IX are very high.

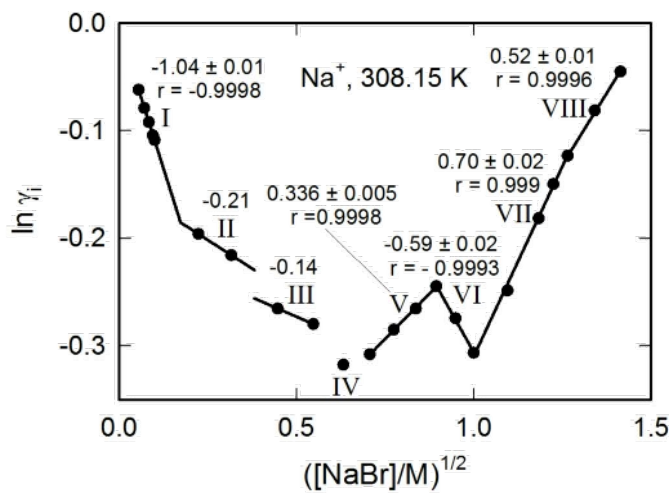

**Fig. 3.** Eight phases. Transitions at 0.173, between 0.316 and 0.447 (jump), between 0.548 and 0.632, between 0.632 and 0.707, and at 0.894, 1.000 and 1.265. Insufficiently detailed data for resolution of phase IV. The absolute r values of lines I, V, VI, VII and VIII are high to very high.

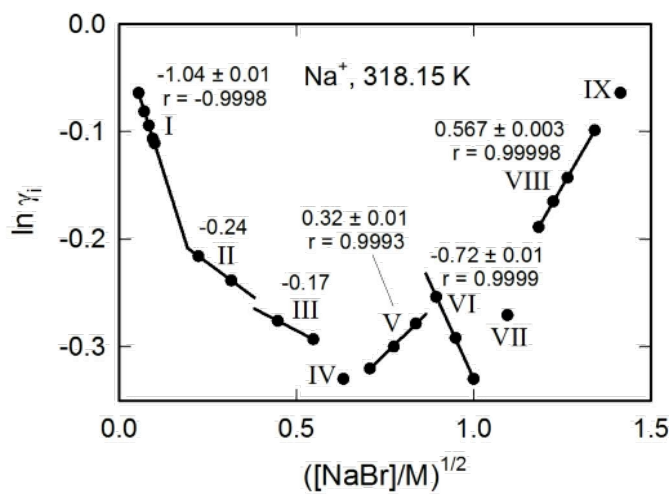

**Fig. 4.** Nine phases. Transitions at 0.193, between 0.316 and 0.447 (jump), between 0.548 and 0.632, between 0.632 and 0.707, between 0.837 and 0.894 (jump), between 1.000 and 1.095, between 1.095 and 1.183, and between 1.342 and 1.414. Insufficiently detailed data for resolution of phases IV, VII and IX. The absolute r values of lines I, V, VI and VIII are very to exceedingly high.

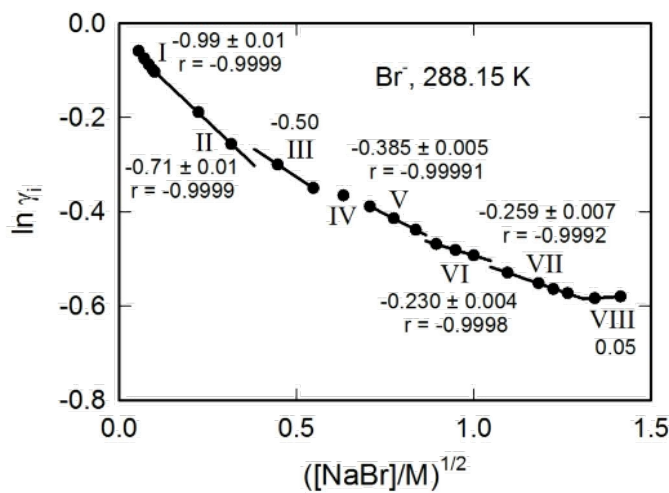

**Fig. 5.** Eight phases. Transitions at 0.100, between 0.316 and 0.447 (jump), between 0.548 and 0.632, between 0.632 and 0.707, between 0.837 and 0.894 (small jump), between 1.000 and 1.095 small (jump), and at 1.310. Insufficiently detailed data for resolution of phase IV. Lines VI and VII are about parallel. The absolute r values of lines I, II, V, VI and VII are very high.

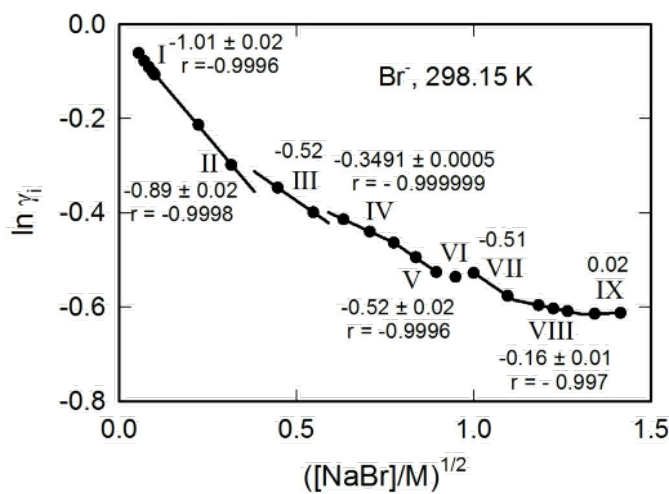

**Fig. 6.** Nine phases. Transitions at 0.100, between 0.316 and 0.447 ( jump), between 0.548 and 0.632 (small jump), at 0.775, between 0.894 and 0.949, between 0.949 and 1.000, at 1.113 and 1.304. Insufficiently detailed data for resolution of phase VI. Lines III, V and VII are parallel. The absolute r values of lines I, II, IV and V are very high.

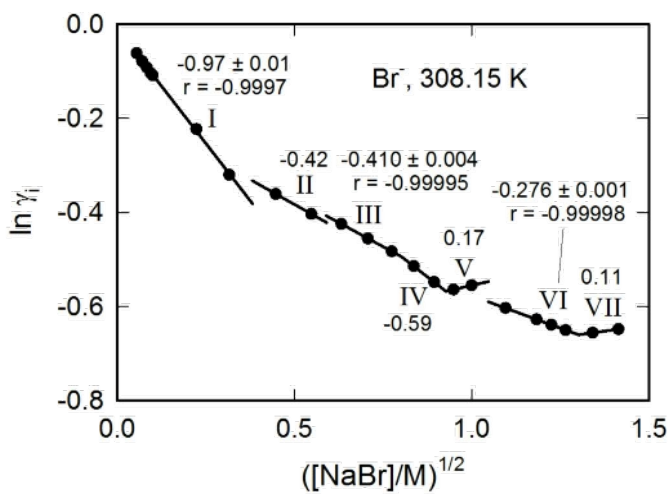

**Fig. 7.** Seven phases. Transitions between 0.316 and 0.447 (jump), between 0.548 and 0.632 (tiny jump), at 0.802 and 0.927, between 1.000 and 1.095 (jump), and at 1.302. Lines II and III are parallel. The absolute r values for lines I, III and VI are very high.

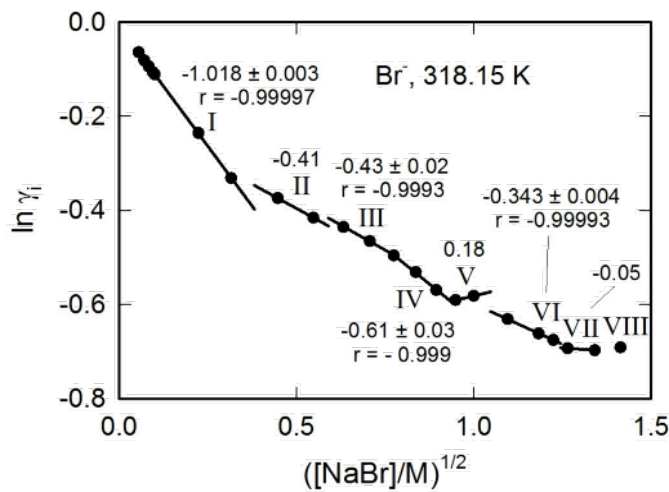

**Fig. 8.** Eight phases. Transitions between 0.316 and 0.447 (jump), between 0.548 and 0.632 (tiny jump), at 0.795 and 0.935, between 1.000 and 1.095 (jump), between 1.225 and 1.265 (tiny jump), and between 1.342 and 1.414. Lines II and III are parallel. The absolute r values for lines I, III, IV and VI are high to very high.

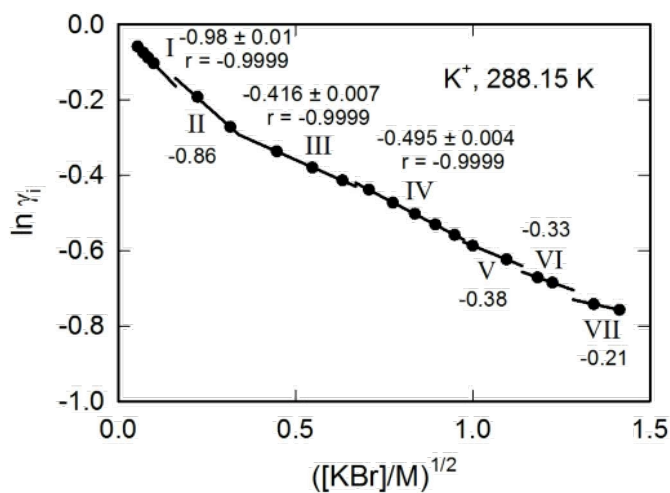

**Fig. 9.** Seven phases. Transitions between 0.100 and 0.224 (tiny jump), at 0.341, between 0.632 and 0.707 (tiny jump), between 0.949 and 1.000 (tiny jump), between 1.095 and 1.183 (tiny jump), and between 1.225 and 1.342 (small jump). Lines V and VI are about parallel. Very high absolute r values for lines I, III and IV.

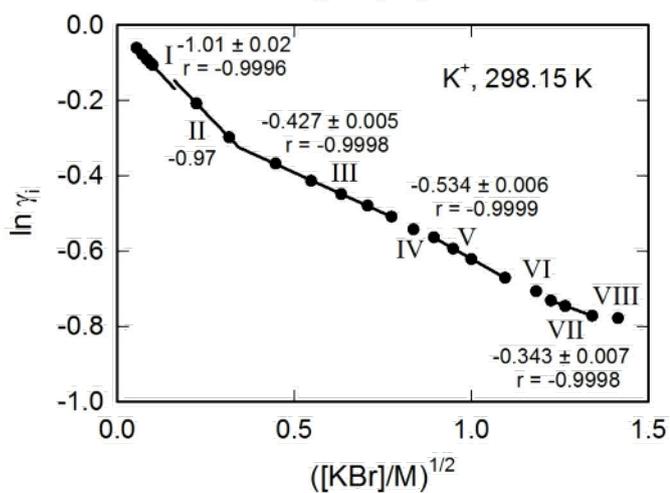

**Fig. 10.** Eight phases. Transitions between 0.100 and 0.224 (tiny jump), at 0.344, between 0.775 and 0.837, between 0.837 and 0.894, between 1.095 and 1.183, between 1.183 and 1.225, and between 1.342 and 1.414. Insufficiently detailed data for resolution of phases IV, VI and VIII. Lines I and II are parallel. Very high absolute r values for lines I, III, V and VII.

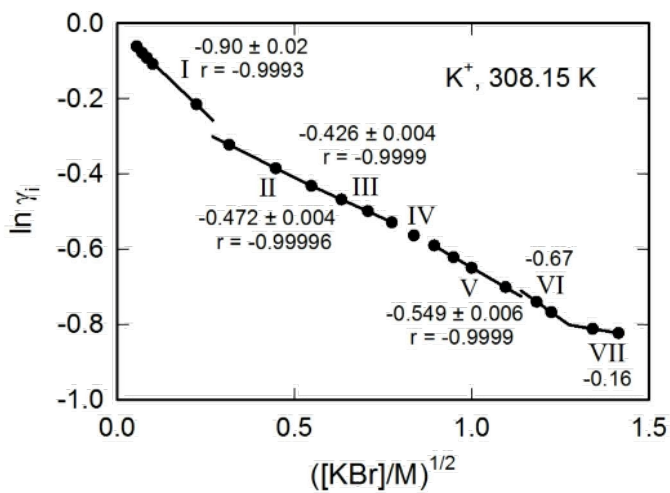

**Fig. 11.** Seven phases. Transitions between 0.224 and 0.316 (jump), at 0.548, between 0.775 and 0.837, between 0.837 and 0.894, between 1.095 and 1.183 (tiny jump), and at 1.275. Insufficiently detailed data for resolution of phase IV. Lines II and III are about parallel. Very high absolute r values for lines I, II, III and V.

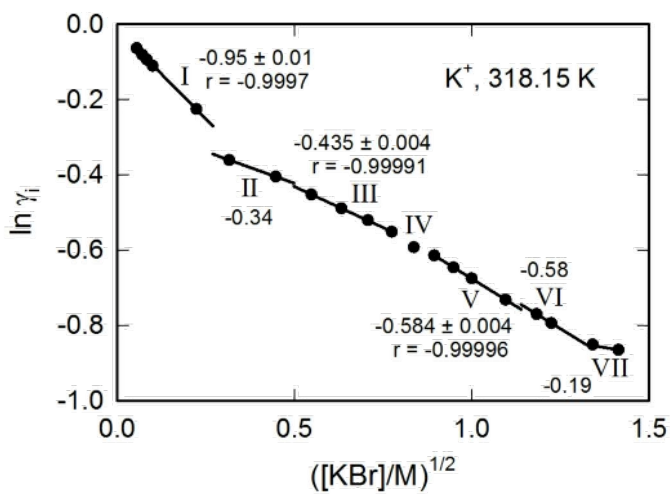

**Fig. 12.** Seven phases. Transitions between 0.224 and 0.316 (jump), between 0.447 and 0.548 (tiny jump), between 0.775 and 0.837, between 0.837 and 0.894, between 1.095 and 1.183 (tiny jump), and at 1.323. Insufficiently detailed data for resolution of phase IV. Lines V and VI are precisely parallel. Very high absolute r values for lines I, III and V.

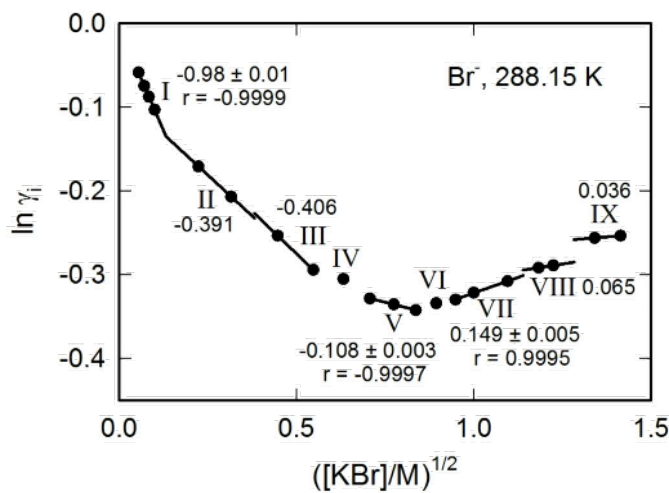

**Fig. 13.** Nine phases. Transitions at 0.132, between 0.316 and 0.447 (tiny jump), between 0.548 and 0.632, between 0.632 and 0.707, between 0.837 and 0.894, between 0.894 and 0.949, between 1.095 and 1.183 (tiny jump), and between 1.225 and 1.342 (jump). Insufficiently detailed data for resolution of phases IV and VI. Lines II and III are parallel as are, roughly, also lines VIII and IX. Very high absolute r values for lines I, V and VII.

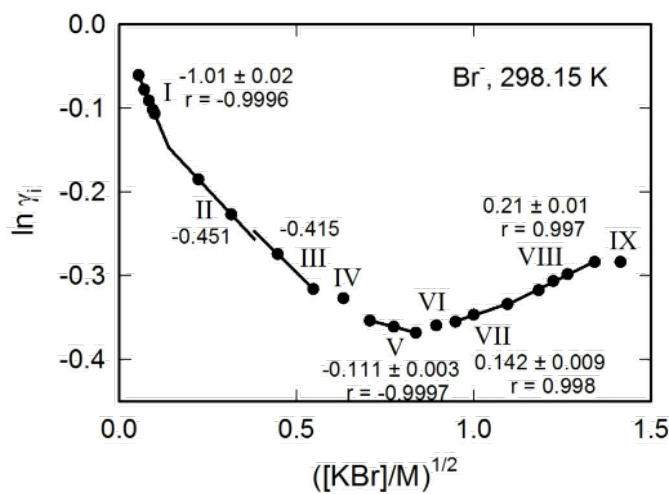

**Fig. 14.** Nine phases. Transitions at 0.140, between 0.316 and 0.447 (tiny jump), between 0.548 and 0.632, between 0.632 and 0.707, between 0.837 and 0.894, between 0.894 and 0.949, at 1.109, and between 1.342 and 1.414. Insufficiently detailed data for resolution of phases IV, VI and IX. Lines II and III are parallel. Very high absolute r values for phases I and V

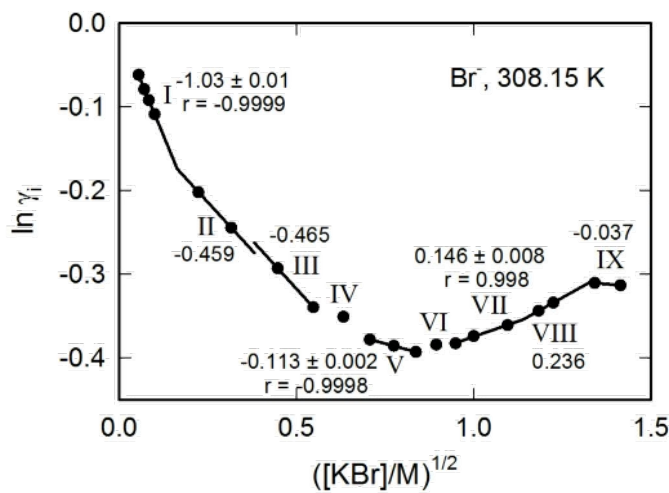

**Fig. 15.** Nine phases. Transitions at 0.164, between 0.316 and 0.447 (tiny jump), between 0.548 and 0.632, between 0.632 and 0.707, between 0.837 and 0.894, between 0.894 and 0.949, and at 1.139 and 1.327. Insufficiently detailed data for resolution of phases IV and VI. Lines II and III are parallel. Very high absolute r values for phases IV and VI.

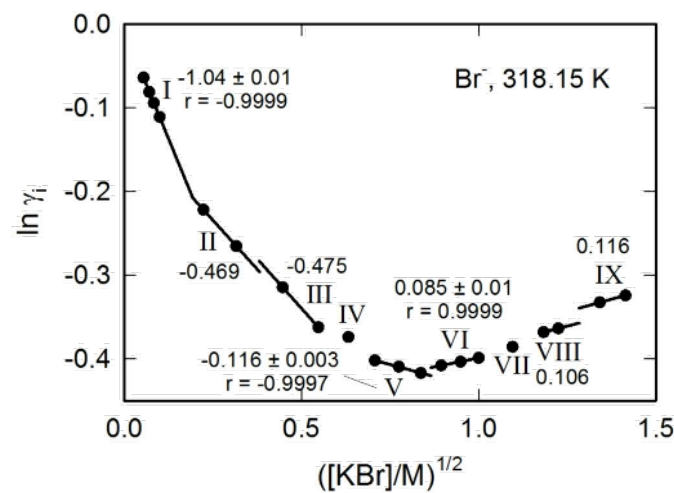

**Fig. 16.** Nine phases. Transitions at 0.193, between 0.316 and 0.447 (tiny jump), between 0.548 and 0.632, between 0.632 and 0.707, between 0.837 and 0.894 (tiny jump), between 1.000 and 1.095, between 1.095 and 1.183, and between 1.225 and 1.342 (jump). Insufficiently detailed data for resolution of phases IV and VII. Lines II and III and, approximately, lines VIII and IX are parallel. Very high absolute r values for phases I, V and VI.

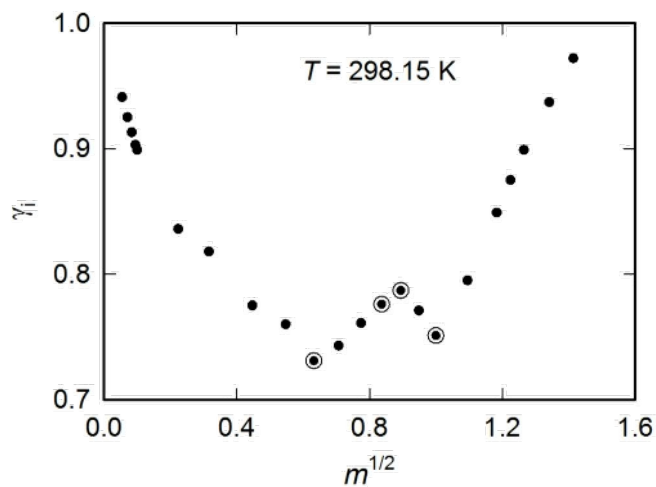

Fig. 17. Data from Table 3 in Lee et al. 2002. The circled points are not included in their Fig. 4.

## Conclusions and Questions

The finding of multiphasic profiles for ion activities has far-reaching implications and raises difficult questions (Nissen 2017). These will not be reiterated here, but complex profiles for activation, such as the present ones by $Na^+$, are spectacular and may prove useful in further studies.

**Acknowledgment –** I am very grateful to Bob Eisenberg for his continued interest and encouragement.